\documentclass[11pt]{article}  
\usepackage{menuproc}
\usepackage{cite}
\usepackage{epsfig}
\usepackage{amsmath,amssymb}
\begin{document}
%
%
%
\titlematter{Recent progress in pion photo- and electroproduction analysis}
{L. Tiator$^a$, D. Drechsel$^a$, S. Kamalov$^b$ and S. N. Yang$^c$}%
{$^a$Institut f\"ur Kernphysik, Universit\"at Mainz, 55099 Mainz,
Germany\\ $^b$Laboratory of Theoretical Physics, JINR, 141980
Dubna, Russia\\ $^c$Department of Physics, National Taiwan
University, Taipei, Taiwan}

{Pion photo- and electroproduction has been studied at threshold
and in the resonance region below $W<2$ GeV. At threshold $\pi^0$
production can be very well explained within a dynamical model
derived from an effective chiral Lagrangian. The final state
interaction is nearly saturated by single charge exchange
rescattering. In the resonance region new electroproduction data
at $Q^2=1$ GeV$^2$ has been analyzed with $MAID$ and longitudinal
and transverse photon helicity amplitudes have been determined for
different resonances. A detailed study of the $E/M$ and $S/M$
ratios of the $N\rightarrow\Delta$ transition shows a zero
crossing of $R_{EM}$ near $Q^2=4$ GeV$^2$, whereas the $R_{SM}$
becomes increasingly negative at large $Q^2$.}

\section{Introduction}

The unitary isobar model
$MAID$ is a model for single pion photo- and electroproduction off
protons and neutrons \cite{MAID}. It is based on a non-resonant
background described by Born terms and vector meson exchange
contributions and nucleon resonance excitations modeled by
Breit-Wigner functions
\begin{equation}
t_{\gamma\pi}^{\alpha}=t_{\gamma\pi}^{B,\alpha}+t_{\gamma\pi}^{R,\alpha}\,.
\end{equation}
Both parts, background and resonance are separately unitarized.
This has been achieved by a K-matrix unitarization in the case of
the background
\begin{equation}
 t_{\gamma\pi}^{B,\alpha}(MAID)=
 \exp{(i\delta_{\alpha})}\,\cos{\delta_{\alpha}}
 v_{\gamma\pi}^{B,\alpha}(W,Q^2)\,,
\end{equation}
and by introducing a unitary phase $\phi_R$ for the resonance
excitations
\begin{equation}
t_{\gamma\pi}^{R,\alpha}(W,Q^2)\,=\,{\bar{\cal
A}}_{\alpha}^R(Q^2)\, \frac{f_{\gamma R}(W)\Gamma_R\,M_R\,f_{\pi
R}(W)}{M_R^2-W^2-iM_R\Gamma_R} \,e^{i\phi_R}\,. \label{eq:BW}
\end{equation}
The phases $\delta_{\alpha}$ are the elastic pion-nucleon
scattering phases in a particular channel $\alpha=\{l,j,t\}$ below
the inelastic threshold of two-pion production. In order to take
account of inelastic effects, the factor $\exp{(i\delta_{\alpha})}
\cos{\delta_{\alpha}}$ is replaced by
$\frac{1}{2}[\eta_{\alpha}\exp{(2i\delta_{\alpha})} +1]$ with the
inelasticity parameters $\eta_{\alpha}$ at higher energies.
Additional background terms are included to account for S- and
P-wave pion loop effects.

In the case of the Dynamical Model ($DMT$) \cite{KY99}, the
background contribution is given by
\begin{equation}
t^{B,\alpha}_{\gamma\pi}(DMT)=e^{i\delta_{\alpha}}\cos{\delta_{\alpha}}
\left[v^{B,\alpha}_{\gamma\pi} + P\int_0^{\infty} dq'
\frac{{q'}^2\,R_{\pi N}^{(\alpha)}(q,q')
\,v^{B,\alpha}_{\gamma\pi}(q')}{W-E_{\pi N}(q')}\right]
\label{eq:Tback}
\end{equation}
with the full $\pi N$ scattering
reaction matrix $R^{(\alpha)}_{\pi N}$. In this case the pion loop
effects that are especially important near threshold are generated
dynamically and show up as a principal value integral over the
reaction matrix.

Presently, we have included 8 nucleon resonances, 3 Deltas:
$P_{33}(1232)$, $S_{31}(1620)$ and $D_{33}(1700)$ and 5 $N^*s$:
$P_{11}(1440)$, $D_{13}(1520)$, $S_{11}(1535)$, $S_{11}(1650)$ and
$F_{15}(1680)$. All of them are included with longitudinal and
transverse electromagnetic couplings. The corresponding helicity
amplitudes $A_{1/2}$, $A_{3/2}$ and $S_{1/2}$ can be fitted to
experimental data and can be freely changed in the $MAID$ program.

We will show results obtained with the Dynamical Model for the
threshold region and present some recent fits to newer
electroproduction data with $MAID$ for the resonance region with a
discussion of the $Q^2$ evolution of the $E/M$ and $S/M$ ratios of
the $N\rightarrow \Delta(1232)$ transition.

\section{Results in the threshold region}

For $\pi^0$ photoproduction, we first calculate the multipole
$E_{0+}$ near threshold by solving the coupled channels equation
within a basis with physical pion and nucleon masses. The coupled
channels equation leads to the following expression for the pion
photoproduction t-matrix in the $\pi^0p$ channel:
\begin{eqnarray}
t_{\gamma\pi^0}(W)& = & v_{\gamma\pi^0}(W)+v_{\gamma\pi^0}(W)\,
g_{\pi^0 p}(W)\,t_{\pi^0 p\rightarrow \pi^0 p}(W) \nonumber\\& + &
v_{\gamma\pi^+}(W)\, g_{\pi^+ n}(W)\,t_{\pi^+ n\rightarrow \pi^0
p}(W)\,, \label{eq:coupled}
\end{eqnarray}
where $t_{\pi^0 p\rightarrow \pi^0 p}$ and $t_{\pi^+ n\rightarrow
\pi^0 p}$ are the $\pi N$  t-matrices in the elastic and charge
exchange channels, respectively. They are obtained by solving the
coupled channels equation for $\pi N$ scattering using the
meson-exchange model of Ref.~\cite{hung}. Our results for
$Re\,E_{0+}$ show that practically all of the final state
interaction effects originate from the $\pi^+ n$ channel and
mainly stems from the principal value integral of Eq.
(\ref{eq:coupled}). In this approach the $t_{\pi N}$ matrix
contains the effect of $\pi N$ rescattering to all orders.
However, we have indeed found that only the first order
rescattering contribution, i.e. the one-loop diagram, is
important. This indicates that the one-loop calculation in ChPT is
a reliable approximation for $\pi^0$ production in the threshold
region.

\begin{figure}[htb]
\centerline{\epsfig{file=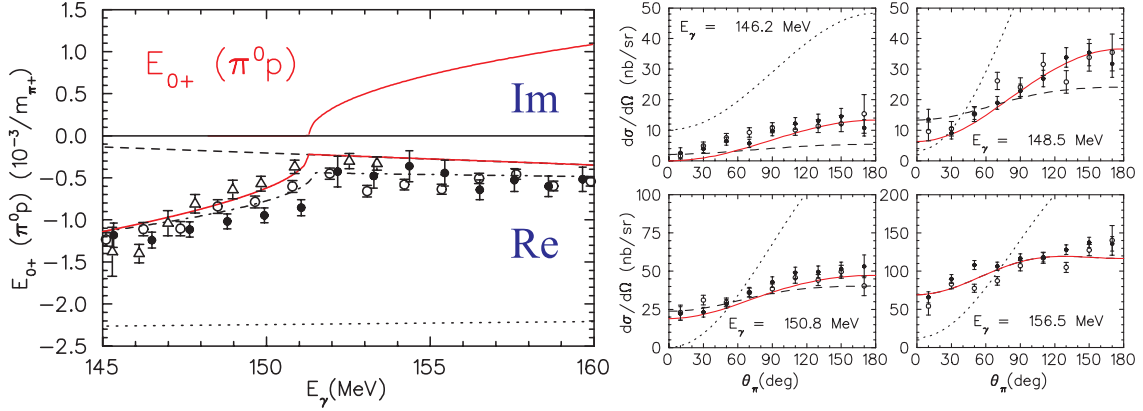,width=15cm,silent=,clip=}}
\caption{\label{fig:thresh1} Real and imaginary parts of the
$E_{0+}$ multipole and differential cross sections below and above
$\pi^+$ threshold for $\gamma p\rightarrow \pi^0 p$. The solid and
dashed curves obtained with and without the cusp effect,
respectively. The dotted curve is without FSI. The dash-dotted
curve is the result of ChPT \protect\cite{Bernard}. Data points
for $E_{0+}$ from Mainz \protect\cite{Fuchs96}($\triangle $),
\protect\cite{Schmidt}($\circ$) and Saskatoon
\protect\cite{Bergstrom}($\bullet$), and for $d\sigma$ from Mainz
\protect\cite{Fuchs96}($\bullet$),
\protect\cite{Schmidt}($\circ$).}
\end{figure}

If the  FSI effects are evaluated with the assumption of isospin
symmetry (IS), i.e., with averaged masses in the free pion-nucleon
propagator, the energy dependence in $Re\, E_{0+}$ in the
threshold region is very smooth. Below $\pi^+$ threshold the
strong energy dependence (cusp effect) \cite{Bernstein98} only
appears because of the pion mass difference and, as we have seen
above, is related to the coupling with the  $\pi^+n$ channel. In
most calculations, the effects from the pion mass difference below
the $\pi^+$ production channel are taken into account by using the
K-matrix approach \cite{Bernard},
\begin{eqnarray}
Re\, E_{0+}^{\gamma\pi^0} = Re\, E_{0+}^{\gamma\pi^0}(IS)
 -  a_{\pi N}\,\omega_c\,Re\,
E_{0+}^{\gamma\pi^+}(IS)\sqrt{1-\frac{\omega^2}{\omega^2_c}}\,,
\label{eq:kmatr}
\end{eqnarray}
where $\omega$ and $\omega_c$ are the $\pi^0$ and $\pi^+$ c.m.
energies corresponding to  $W=E_p + \omega_{\gamma}$ and
$m_n+m_{\pi^+}$, respectively, and $a_{\pi N}=0.124/m_{\pi^+}$ is
the pion charge exchange threshold amplitude. $E_{0+}^{\gamma
\pi^{0,+}}(IS)$ is the $\pi^{0,+}$ photoproduction amplitude
obtained with the assumption of isospin symmetry (IS), i.e.,
without the pion mass difference in Eq. (\ref{eq:Tback}). Such an
approximation is often used in the data analysis in order to
parametrize the $E_{0+}$ multipole below $\pi^+n$ threshold in the
form of $E_{0+}(E)=a + b \sqrt{1-(\omega/\omega_c)^2}$.
Numerically this approximation is very precise and differs only by
10\% at the $\pi^0$ threshold and becomes indistinguishable above
the $\pi^+$ threshold.  In Fig. \ref{fig:thresh1} the results
obtained within this approximation scheme are represented by the
solid curve and compared to the ChPT calculation (dash-dotted
curve) \cite{Bernard91}. Over the whole energy range the
difference is rather small and within the experimental
uncertainties. Huge effects, however, arise if the cusp would be
neglected or obviously  if the FSI effects would be totally
ignored (dotted curve).

In  Fig. \ref{fig:thresh1} we also compare the predictions of our
model for the differential cross section with recent
photoproduction data from Mainz \cite{Fuchs96,Schmidt}. The dotted
and solid curves are obtained without and with FSI effects,
respectively. It is seen that both off-shell pion rescattering and
cusp effect substantially improve the agreement with the data.
This indicates that our model gives reliable predictions also for
the threshold behaviour of the $P$-waves without any additional
arbitrary parameters.

\begin{figure}[htb]
\centerline{\epsfig{file=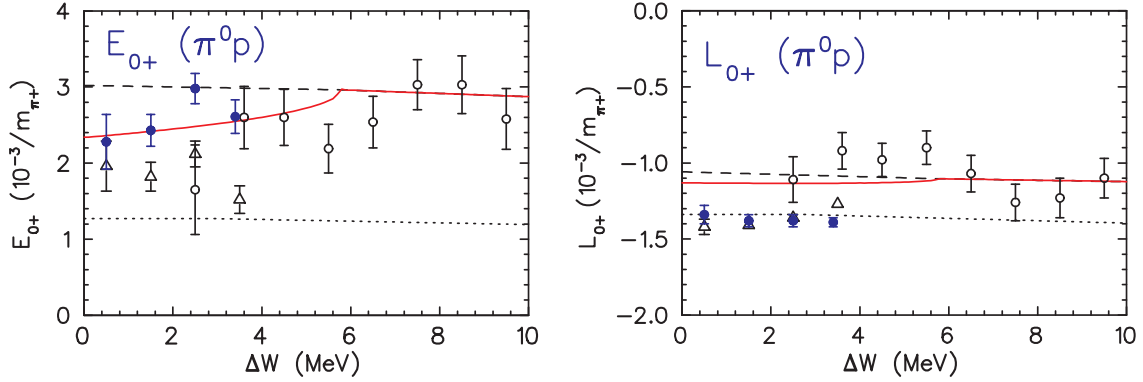,width=15cm,silent=,clip=}}
\caption{\label{fig:thresh2} Real parts of $E_{0+}$ and $L_{0+}$
for $e p \rightarrow e'\pi^0 p$ at $Q^2$=0.1 (GeV/c)$^2$.
Notations are the same as in Fig. 1. Data points from
NIKHEF\protect\cite{NIKHEF}($\circ$) and
Mainz\protect\cite{Distler}($\triangle$). The results of the
present work obtained by using the $P$-waves of our model are
given by ($\bullet$).}
\end{figure}

Pion electroproduction provides us with information on the $Q^2 =
-k^2$ dependence of the transverse $E_{0+}$ and longitudinal
$L_{0+}$ multipoles in the threshold region. The "cusp" effects in
the $L_{0+}$ multipole is taken into account in a similar way as
in the case of $E_{0+}$,
\begin{eqnarray}
Re\, L_{0+}^{\gamma\pi^0} = Re\, L_{0+}^{\gamma\pi^0}(IS) - a_{\pi
N} \,\omega_c\,Re\,
L_{0+}^{\gamma\pi^+}(IS)\,\sqrt{1-\frac{\omega^2}{\omega^2_c}}\,,
\label{eq:L-Kmatrix}
\end{eqnarray}
where all the multipoles are functions of total c.m. energy $W$
and virtual photon four-momentum squared $Q^2$. It is known that
at threshold, the $Q^2$ dependence is given mainly by the Born
plus vector meson contributions in $v_{\gamma\pi}^B$, as described
in Ref.~\cite{MAID}. Similar to pion photoproduction, the K-matrix
approximation and full calculation agree with each other within a
few percent. In Fig. 2 we show our results for the cusp and FSI
effects in the $E_{0+}$ and $L_{0+}$ multipoles for $\pi^0$
electroproduction at $Q^2=0.1$ (GeV/c)$^2$, along with the results
of the multipole analysis from NIKHEF\cite{NIKHEF} and
Mainz\cite{Distler}. Note that results of both groups were
obtained using the $P$-wave predictions given by ChPT. However,
there exist substantial differences between the $P-$wave
predictions of ChPT and our model at finite $Q^2$. To understand
the consequence of these differences, we have made a new analysis
of the Mainz data\cite{Distler} for the differential cross
sections, using our $DMT$ prediction for the $P$-wave multipoles
instead. The $S$-wave multipoles extracted this way are also shown
in Fig. \ref{fig:thresh2} by solid circles. We see that the
results of such a new analysis gives $E_{0+}$ multipoles closer to
the NIKHEF data and in better agreement with our dynamical model
prediction. However, the results of our new analysis for the
longitudinal $L_{0+}$ multipoles stay practically unchanged from
the values found in the previous analyses. Note that the dynamical
model prediction for $L_{0+}$ again agrees much better with the
NIKHEF data. Further details are given in Ref.\cite{Kama02}

In contrast to $DMT$, in $MAID$ the FSI effects are taken into
account using the K-matrix approximation, namely without the
inclusion of off-shell pion rescattering contributions (principal
value integral) in Eq. (\ref{eq:Tback}). As a result, the $S$-,
$P$-, $D$- and $F$-waves of the background contributions are
defined as
\begin{eqnarray}
t_{\alpha}^B(MAID)=\exp{(i\delta_{\alpha})}\,\cos{\delta_{\alpha}}
\,v_{\alpha}^B(q_E,k). \label{eq:Tmaid}
\end{eqnarray}
However, as we have found above, dynamical model calculations show
that pion off-shell rescattering is very important at low pion
energies. The prediction of $MAID$ for $E_{0+}(\pi^0 p)$ at
threshold, represented by the dotted curves in Fig.
\ref{fig:thresh1} lies substantially below the data. It turns out
that it is possible to improve $MAID$, in the case of $\pi^0$
production at low energies, by introducing a phenomenological term
and including the cusp effect of Eq. (\ref{eq:kmatr}). In this
extended version of $MAID2000$, we write the $E_{0+}(\pi^0p)$
multipole as
\begin{equation}
Re\, E_{0+}^{\gamma\pi^0}  =  Re\, E_{0+}^{\gamma\pi^0}(MAID98) +
E_{cusp}(W,Q^2) + E_{corr}(W,Q^2)\,, \label{eq:Emaid}
\end{equation}
where
\begin{equation}
E_{cusp}(W,Q^2)= - a_{\pi N} \,\omega_c\,Re\,
E_{0+}^{\gamma\pi^+}(MAID98)\,\sqrt{1-\frac{\omega^2}{\omega^2_c}}\,.
\label{eq:cusp}
\end{equation}
The phenomenological term $E_{corr}$ which emulates the pion
off-shell rescattering corrections (or pion-loop contribution in
ChPT) can be parameterized in the form
\begin{equation}
E_{corr}(W,Q^2)= \frac{A}{(1 + B^2q^2_{\pi})^2}\,F_D(Q^2)\,,
\label{eq:corr}
\end{equation}
where $F_D$ is the standard nucleon dipole form factor. The
parameters $A$ and $B$ are obtained by fitting to the low energy
$\pi^0$ photoproduction data: $A= 2.01 \times 10^{-3}/m_{\pi^+}$
and $B=0.71 \, fm$.

\section{Results in the resonance region}
\begin{figure}[htb]
\centerline{\epsfig{file=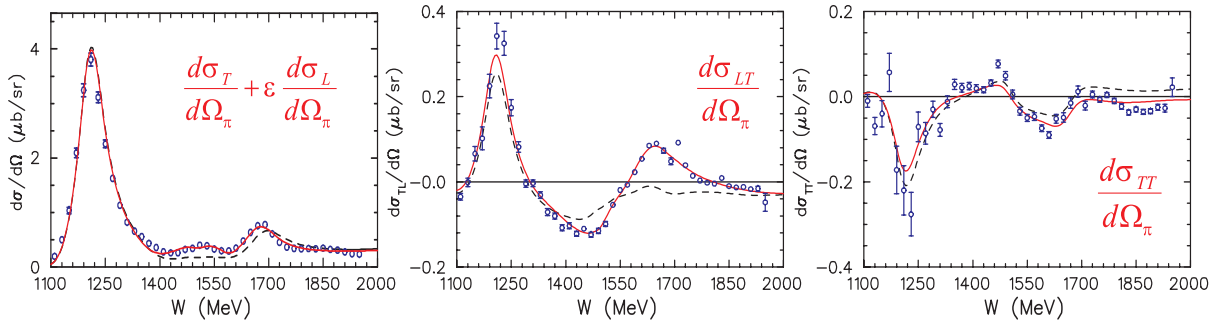,width=16cm,height=4.2cm,silent=,clip=}}
\caption{\label{fig:laveiss} Preliminary experimental results from
the JLab Hall A collaboration\protect\cite{laveiss} at $Q^2=1.0$
GeV$^2$, $\theta_\pi=167^0$ and $\epsilon=0.9$ as functions of the
c.m. energy W. The dashed lines show the standard $MAID2000$
calculations and the solid curves are the results of the fit to
the data.}
\end{figure}

For pion photoproduction in the resonance region we have recently
performed a fit of $MAID$ for both the low-energy $\Delta$ region
and the medium-energy resonance region up to $W=1700$ MeV, where
resonance parameters have been obtained as a part of the
BRAG\footnote{Baryon Resonance Analysis Group,
\texttt{http://cnr2.kent.edu/$\sim$manley/BRAG.html}} partial wave
benchmark analysis\cite{benchmark}. In Fig. \ref{fig:laveiss} we
present a new fit of preliminary results on electroproduction,
$p(e,e'p)\pi^0$, measured by the JLab Hall A
collaboration\cite{laveiss}. The data has been taken at backward
angles at $Q^2=1.0$ GeV$^2$ in the c.m. energy range from $0.95$
GeV to $2.0$ GeV. With a dataset of 363 data points in 3
observables, $d\sigma=d\sigma_T + \epsilon d\sigma_L,
d\sigma_{LT}$ and $d\sigma_{TT}$ and pion angles of $146, 151$ and
$167$ degrees we performed a data analysis with $MAID$.

From the 20 possible resonance parameters we have varied 18 by
fixing $E_{2-}$ and $S_{2-}$ of the $D_{13}(1520)$ because we did
not find enough sensitivity in the data which are only taken at
backward angles. In table \ref{resonances} we give the result of
our fit both for the multipole and the helicity amplitudes. The
multipole amplitudes are compared to the default values of $MAID$.
For the $\Delta(1232)$ resonance we give in addition the $E/M$ and
the $S/M$ ratios. Both are consistent with the previous $MAID$
fits to photo- and electroproduction \cite{Kama01}. The $R_{SM}$
ratio is very well determined by the $d\sigma_{LT}$ data and shows
the tendency to larger negative values for increasing $Q^2$, while
the $R_{EM}$ ratio is much more uncertain and also the model
uncertainties are larger than for the $S/M$ ratio. From
$d\sigma_{LT}$ we also find a large sensitivity to the $S_{0+}$
amplitude of the $S_{11}(1535)$ resonance in the minimum around
$W=1500$ MeV as well as for the $S_{2-}$ amplitude of the
$D_{33}(1700)$ resonance in the second maximum around $W=1650´$
MeV. Furthermore most of the structure in $d\sigma$ and in
$d\sigma_{TT}$ above $W=1700$ MeV is explained by the $M_{2-}$
amplitude of the $D_{33}(1700)$ resonance. However, the $MAID$
model does not include higher resonances so far.

In order to get information on resonance properties a careful
analysis has to be taken. First of all, a partial wave
decomposition in terms of multipoles is necessary to get
information on quantum numbers as angular momentum, spin, parity
and isospin. Second, a background separation is needed, as
especially in pion production a large background is produced by
the strong pion nucleon coupling. However, nucleon Born terms and
vector meson exchange contributions are not the only source of
background. Loop effects can give very large contributions
especially for S- and P-waves, the most famous example is the
$E_{0+}$ in $\pi^0$ photoproduction at threshold. Our resonance
extraction is based on the imaginary parts of the full multipoles
in a specific spin-isospin channel at the resonance position. This
minimizes the model dependence since resonance positions are
mostly well known and this method can be applied to any given
partial wave analysis. It is very similar to the method that has
been applied for the $\Delta(1232)$, however, as Watson's theorem
is no longer fulfilled for higher resonances, some uncertainties
have to be accepted. The helicity amplitudes extracted by this way
are ``dressed'' amplitudes and contain contributions from vertex
corrections. They could be undressed by subtracting background
contributions and unitarization corrections. By such a procedure
they could be directly related to the Breit-Wigner couplings
${\bar{\cal A}}_{\alpha}^R(Q^2)$ of Eq. (\ref{eq:BW}), however,
such a method will always be model dependent.

The helicity amplitudes $A_{1/2}, A_{3/2}$ and $S_{1/2}$ are
determined from the pion electroproduction multipoles at the
resonance position
\begin{eqnarray}
A_{1/2}^{\ell+}   &=& -\frac{1}{2ac_I}
 [(\ell+2)\tilde{E}_{\ell+} +\ell\tilde{M}_{\ell+}]\\
A_{1/2}^{(\ell+1)-} &=& +\frac{1}{2ac_I}
 [(\ell+2)\tilde{M}_{(\ell+1)-} -\ell\tilde{E}_{(\ell+1)-}]\\
A_{3/2}^{\ell+}   &=& +\frac{1}{2ac_I}\sqrt{\ell(\ell+2)}
 [\tilde{E}_{\ell+} -\tilde{M}_{\ell+}]\\
A_{3/2}^{(\ell+1)-} &=& -\frac{1}{2ac_I}\sqrt{\ell(\ell+2)}
 [\tilde{E}_{(\ell+1)-} +\tilde{M}_{(\ell+1)-}]\\
S_{1/2}^{\ell+}   &=& -\frac{1}{\sqrt{2}ac_I}
 (\ell+1)\tilde{S}_{\ell+}\\
S_{1/2}^{(\ell+1)-} &=& -\frac{1}{\sqrt{2}ac_I}
 (\ell+1)\tilde{S}_{(\ell+1)-}
\end{eqnarray}
\begin{equation}
\mbox{with}\quad
a=\sqrt{\frac{1}{\pi}\frac{k_W^R}{q_\pi^R}\frac{1}{2J+1}
\frac{m_N}{M_R}\frac{\Gamma_\pi}{\Gamma_{tot}^2}}\quad
\mbox{and}\quad c_I=\left\{ \begin{array}{r@{\quad:\quad}l}
                        -\sqrt{1/3} & I=1/2\\
                         \sqrt{3/2} & I=3/2
                         \end{array} \right.\,.
\end{equation}
The equivalent photon energy $k_W^R$ and the pion momentum
$q_\pi^R$ are given in the c.m. frame and evaluated at the
resonance position, where also the pion electroproduction
multipoles are obtained, $\tilde{A}\equiv \mbox{Im}A(W=M_R)$ for
$A=E,M,S$. For the transverse amplitudes these formulas agree with
PDG and Ref.~\cite{Arnd90}. For longitudinal amplitudes we found
different definitions in the literature, here we use a definition
consistent with notations used in DIS\cite{Abe98}.
\begin{figure}[htb]
\centerline{\epsfig{file=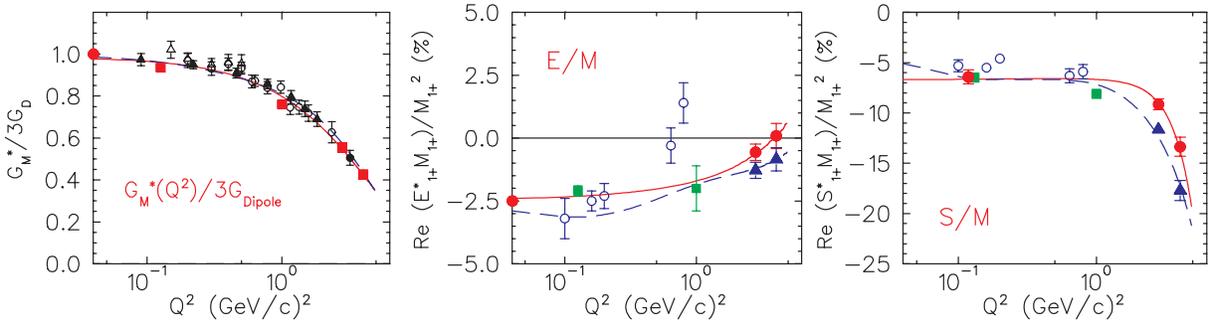,width=16cm,silent=,clip=}}
\caption{\label{fig:delta} The $Q^2$ dependence of the magnetic
$G_M^*$ form factor and the $E/M$ and $S/M$ ratios at $W=1232$
MeV. The solid and dashed curves are the $MAID$ and dynamical
model results, respectively. JLab results discussed here are shown
at $Q^2=1$ (GeV/c)$^2$, Bates results at 0.126 (squares), the
Mainz double polarization result of $S/M$ at 0.12 (full circle),
data at 2.8 and 4.0 from Ref.\protect\cite{Frolov99} (squares) and
preliminary results of the ratios in the range of 0.1 - 0.8 are
from Bonn\protect\cite{Gothe00} (open circles). Results of our
analysis for the ratios at 2.8 and 4.0 are obtained using $MAID$
(full circles) and the dynamical models (triangles). In the case
of $G_M^*$ they fully agree with Ref.\protect\cite{Frolov99}. For
older data of $G_M^*$ see Ref. \cite{Kama01}. The photoproduction
results of Mainz\protect\cite{Beck97} are placed at the lowest
$Q^2$ value. All numbers are given in units of (GeV/c)$^2$.}
\end{figure}

\begin{table}[htb]
\begin{center}
{\footnotesize
\begin{tabular}{|cc|c|c|cc|}
\hline $N^*$ &  &  $MAID$ & $MAID$    &  & helicity   \\
      &  &  defaults    & fit results & & amplitudes \\
\hline $P_{33}(1232)$
   & $\tilde{E}_{1+}^{3/2}$ & -0.44 &  -0.38$\pm$ 0.17 & $A_{1/2}$ &  -72$\pm$ 3 \\
   & $\tilde{M}_{1+}^{3/2}$ &  20.1 & 19.0$\pm$ 0.4 & $A_{3/2}$ & -135$\pm$ 3 \\
   & $\tilde{S}_{1+}^{3/2}$ &  -1.31&  -1.55$\pm$ 0.05 & $S_{1/2}$ &   18$\pm$ 1 \\
   & $R_{EM}$ & -2.2 & -2.0$\pm$ 0.9 &         &                 \\
   & $R_{SM}$ & -6.5 & -8.1$\pm$ 0.2 &         &                 \\
\hline $P_{11}(1440)$
   & $\tilde{M}_{1-}^{1/2}$ & 1.87 &  2.2$\pm$ 0.2 & $A_{1/2}$ &  -59$\pm$ 9 \\
   & $\tilde{S}_{1-}^{1/2}$ & 1.26 &  0.6$\pm$ 0.1 & $S_{1/2}$ &  -11$\pm$ 4 \\
\hline $D_{13}(1520)$
   & $\tilde{E}_{2-}^{1/2}$ & -0.05 &  -0.05         & $A_{1/2}$ &  -47$\pm$ 6 \\
   & $\tilde{M}_{2-}^{1/2}$ & 1.71 & 1.2$\pm$ 0.2 & $A_{3/2}$ &   26$\pm$ 4 \\
   & $\tilde{S}_{2-}^{1/2}$ &    0  &    0          & $S_{1/2}$ &         0       \\
\hline $S_{11}(1535)$
   & $\tilde{E}_{0+}^{1/2}$ & 4.37 &  3.5$\pm$ 0.8 & $A_{1/2}$ &   61$\pm$ 14 \\
   & $\tilde{S}_{0+}^{1/2}$ & 0.1 & -1.2$\pm$ 0.1 & $S_{1/2}$ &  -15$\pm$ 2 \\
\hline $S_{31}(1620)$
   & $\tilde{E}_{0+}^{3/2}$ & 0.21   & 3.4$\pm$ 0.3 & $A_{1/2}$ &  -50$\pm$5 \\
   & $\tilde{S}_{0+}^{3/2}$ & 0.89   & 1.7$\pm$ 0.5 & $S_{1/2}$ &  -18$\pm$ 7 \\
\hline $S_{11}(1650)$
   & $\tilde{E}_{0+}^{1/2}$ & 2.8 &  3.0$\pm$ 0.4 & $A_{1/2}$ &   43$\pm$ 6 \\
   & $\tilde{S}_{0+}^{1/2}$ &  0  & -1.2$\pm$ 0.6 & $S_{1/2}$ &  -12$\pm$ 6 \\
\hline $F_{15}(1680)$
   & $\tilde{E}_{3-}^{1/2}$ & 0.32 & -0.06$\pm$ 0.03 & $A_{1/2}$ &  -52$\pm$ 9 \\
   & $\tilde{M}_{3-}^{1/2}$ & 0.83 & 0.80$\pm$ 0.05 & $A_{3/2}$ &   33$\pm$ 9 \\
   & $\tilde{S}_{3-}^{1/2}$ &    0  & -0.10$\pm$ 0.03 & $S_{1/2}$ &  -7$\pm$ 2 \\
\hline $D_{33}(1700)$
   & $\tilde{E}_{2-}^{3/2}$ & -0.85 & -1.0$\pm$ 0.2 & $A_{1/2}$ &  104$\pm$12 \\
   & $\tilde{M}_{2-}^{3/2}$ & 0.30 & 1.1$\pm$ 0.1 & $A_{3/2}$ &   -4$\pm$12 \\
   & $\tilde{S}_{2-}^{3/2}$ &    0  & 0.2$\pm$ 0.1& $S_{1/2}$ &  -14$\pm$ 7 \\
\hline PV-PS mixing:
   & $\Lambda_m$ &450& 350 $\pm$ 35 & &\\
\hline
\end{tabular}
}
\end{center}
\caption{Proton resonance multipoles ($\tilde{A}\equiv
\mbox{Im}A(W=M_r)$ in $10^{-3}/m_\pi$), helicity amplitudes (in
$10^{-3}$ GeV$^{-1/2}$) and values of the PV-PS mixing parameter
$\Lambda_m$ (in MeV) as in $MAID2000$ and obtained in our fit at
$Q^2=1.0$ GeV$^2$. The $D_{13}(1520)$ $\tilde{E}_{2-}$ and
$\tilde{S}_{2-}$ amplitudes were fixed. The $E/M$ and $S/M$ ratios
of the $\Delta(1232)$ are given in percentage.} \label{resonances}
\end{table}

In Fig. \ref{fig:delta} we show our extracted values for the
magnetic form factor $G_M^*/3G_D$ and the ratios $R_{EM}$ and
$R_{SM}$ together with other data determined in different ways in
recent experiments and data analyses on a semi-log scale. For
photoproduction the Mainz results\cite{Beck97} are shown, around
0.125 GeV$^2$ the values of the Bates analysis and the result of
the Mainz measurement with recoil polarization, in the medium
$Q^2$ range the preliminary data of Bonn\cite{Gothe00} and at high
$Q^2$ our analysis of the JLab Hall C data\cite{Frolov99} with
$MAID$ and the $DMT$. The main difference between our results and
those of Ref.~\cite{Frolov99} is that our values of $R_{EM}$ show
a clear tendency to cross zero and change sign as $Q^2$ increases.
This is in contrast with the results obtained in the original
analysis \cite{Frolov99} of the data which concluded that $R_{EM}$
would stay negative and tend toward more negative values with
increasing $Q^2$. Furthermore, we find that the absolute value of
$R_{SM}$ is strongly increasing.

\section{Summary}
With the unitary isobar model $MAID$ and the dynamical model $DMT$
we have two very good tools available to analyze data of pion
photo- and electroproduction and to plan new experiments with
increased sensitivity to specific questions. While $DMT$ includes
pion loop contributions that are especially important for low
partial waves (S and P), $MAID$ originally was constructed only
from tree diagrams, resonance excitations and K-matrix
unitarization contributions. Therefore, to get better agreement
for S-waves, $MAID$ has been extended phenomenologically by
low-energy corrections and the unitary cusp effect. At higher
energies the PS-PV mixing of $MAID$ that was already introduced
from the beginning, also serves for this purpose to effectively
taking into account of loop contributions.

With all parameters already fixed by $\pi N$ scattering and pion
photoproduction in the resonance region, $DMT$ describes pion
photoproduction at threshold very well, similar to the
calculations in ChPT or with dispersion relations. For
electroproduction at threshold we also find good agreement with
the experiment at $Q^2=0.1$ GeV$^2$, however, we also have
problems describing the recent Mainz data at $Q^2=0.05$ GeV$^2$ as
it also appears in ChPT calculations. In the resonance region both
models $DMT$ and $MAID$ can equally well describe photo- and
electroproduction data up to $W=1700$ MeV by fitting the photon
helicity amplitudes $A_{1/2}, A_{3/2}$ and $S_{1/2}$ of the
individual resonances. Here, we have demonstrated this with the
preliminary Hall A data of JLab at $Q^2=1.0$ GeV$^2$. Even with a
dataset limited to backward pion angles we are able to determine
quite a few resonance parameters in satisfactory precision and can
also give longitudinal couplings where previous information
practically did not exist. For the $E/M$ and $S/M$ ratios of the
Delta resonance we combine our previous fits of Mainz, Bates and
JLab Hall C data with our new analysis and find a consistent $Q^2$
evolution of these ratios with a slowly rising $R_{EM}$ that
crosses zero around $Q^2=4$ GeV$^2$, and for the longitudinal
coupling a $R_{SM}$ that significantly increases to larger
negative values at high $Q^2$. If expectations from pQCD will be
fulfilled, a sharp rise in the $E/M$ ratio towards 100\% should be
seen in the next generation of experiments above $Q^2=5$ GeV$^2$
and a leveling of the $S/M$ ratio to a constant value.

\acknowledgments{We gratefully acknowledge financial support of
this work in parts by the National Science Council of ROC under
Grant No.~NSC89-2112-M002-078, by Deutsche Forschungsgemeinschaft
(SFB 443), and by a joint project NSC/DFG TAI-113/10/0.}


\end{document}